\documentclass[conference]{IEEEtran}
\IEEEoverridecommandlockouts

\usepackage[cmex10]{amsmath}
\usepackage{amsthm, amssymb, amsfonts}
\usepackage{bbm}

\usepackage{ifthen}
\usepackage{geometry} 

\counterwithin{equation}{section}
\newcounter{thm}
\newtheorem{theorem}[thm]{Theorem}

\newtheorem{definition}[thm]{Definition}

\newtheorem{lemma}[thm]{Lemma}

\newcommand*{\rom}[1]{\uppercase\expandafter{\romannumeral #1\relax}}

\newcommand{\calS}{\mathcal{S}}

\newcommand{\bGamma}{\mathbf{\Gamma}}

\newcommand{\hr}[1]{
    \ifthenelse{\equal{#1}{\string 1}}
  {
   \mathcal{H}%
  }
  {
   \mathcal{H}^{\otimes #1}%
  }
}
\newcommand{\eps}{{\varepsilon}}        

\newcommand{\eins}{{\mathbbm{1}}}
\newcommand{\tr}{{\mathrm{tr}}}
\newcommand{\bm}{\mathbf{m}}

\allowdisplaybreaks

\makeatletter
\newcommand*{\algrule}[1][\algorithmicindent]{%
  \makebox[#1][l]{%
    \hspace*{.2em}
    \vrule height .75\baselineskip depth .25\baselineskip
  }
}

\newcount\ALG@printindent@tempcnta
\def\ALG@printindent{%
    \ifnum \theALG@nested>0
    \ifx\ALG@text\ALG@x@notext
    \else
    \unskip
    \ALG@printindent@tempcnta=1
    \loop
    \algrule[\csname ALG@ind@\the\ALG@printindent@tempcnta\endcsname]%
    \advance \ALG@printindent@tempcnta 1
    \ifnum \ALG@printindent@tempcnta<\numexpr\theALG@nested+1\relax
    \repeat
    \fi
    \fi
}

\newcounter{protocol}

\title{Data Transmission over a Bosonic Arbitrarily Varying Quantum Channel}

\author{
    \IEEEauthorblockN{Janis N\"otzel, Florian Seitz}
    \thanks{This work was financed by the Federal Ministry of Education and Research of Germany via grants 16KISQ039 and 16KISQ077 as well as in the programme of ``Souver\"an. Digital. Vernetzt.''. Joint project 6G-life, project identification number: 16KISK002, and by the DFG via grant NO 1129/2-1. Inspiring discussions with Prakash Narayan are acknowledged.}
    \IEEEauthorblockA{
        Emmy-Noether Gruppe Theoretisches Quantensystemdesign\\
        Technische Universit\"at M\"unchen\\
        janis.noetzel@tum.de, flo.seitz@tum.de
    }
}

\begin{document}

\maketitle
\begin{abstract}
    Arbitrarily varying channels offer a powerful framework for analyzing the robustness of quantum communication systems, especially for classical-quantum models, where the analysis displays strengths or weaknesses of specific signal constellations under generic attacks. In this work, we provide a coding theorem for a large class of practically relevant arbitrarily varying channel models. Namely, we give an explicit capacity formula for the lossy bosonic channel subject to semi-classical attacks, where an adversary injects semi-classical states into the transmission line. Mathematically, this is modeled via a beam-splitter setup, with transmitter and jammer controlling different input ports and the receiver observing one output port.
    We show how a recently conjectured new quantum entropy power inequality relates to our capacity formula.
\end{abstract}
\section{Introduction}

Arbitrarily Varying Channels (AVCs) provide a theoretical model for communication systems operating under hostile interference or jamming and were introduced in \cite{bbt}. The system model turned out to be a fertile ground for further study \cite{ericsonAVC,elimination,csiszarNarayanPositivity}, as the question of whether the capacity was positive or not is nontrivial. The quantum analogue of such channels where data is transmitted was initially examined in~\cite{ahlswede-blinovsky}. Follow-up work \cite{quantumAVC} studied in addition the transmission of entanglement over such channels. Unlike memoryless or compound channels, AVCs present a more intricate structure: the capacity for reliable communication can jump from zero to a strictly positive value when the sender and receiver share common randomness (CR), provided that the channel is not symmetrizable---a concept explored in~\cite{ahlswede-elimination,csiszarNarayanPositivity}. Intriguingly, even classical AVCs can display the phenomenon of superactivation (which was first discovered for quantum systems \cite{Smith_2008}), when secrecy constraints are introduced, as shown in~\cite{schaeferBocheSuperActivation}. The work \cite{noetzelBookchapter} points out relations of the information-theoretic concept of the AVC with an additional eavesdropper to signal processing techniques such as spread spectrum communications. 

While for classical systems, a large body of literature is available that covers the important additive white Gaussian noise case \cite{csiszarNarayan}, much less is known for the corresponding (bosonic) quantum systems. This work investigates the communication capacity $C$ of a fundamental bosonic channel under adversarial conditions. Specifically, we study a beam splitter channel denoted by $\boxplus_\tau$, where the sender controls one input, and a jammer manipulates the other. The receiver accesses a single output mode, with the beam splitter configured such that inputting a vacuum state from the jammer reduces the channel to a pure-loss bosonic model with transmissivity $\tau$. This setup effectively models scenarios such as free-space optical communication or non-amplified fiber links. We restrict the sender to use coherent states and the adversary to use semi-classical states as inputs and establish a full coding theorem for CR-assisted schemes.

Interestingly, the analysis connects with entropy power inequalities (EPIs). In~\cite{csiszarNarayan}, the capacity for a Gaussian AVC with additive power-limited jamming was fully characterized, based on a clever decoding technique which does not translate directly to quantum systems such as the one studied in this work. However, it may be assumed that for these quantum systems the CR-assisted capacity will equal the non-assisted one in case the latter is positive. It is therefore of interest to derive at least a formula for the CR-assisted capacity. A reasonable assumption is that one may prove a capacity formula in the form of a convex optimization problem involving entropic quantities. In the classical model \cite{csiszarNarayan} with zero additive noise but active jamming, such approach would lead to a situation where a Gaussian input $ X $ with zero mean and variance $\sigma^2$ is given, and the CR-assisted capacity takes the form $\min_Y(H(X+Y) - H(Y))$, where $Y$ represents the adversarial noise which is constrained in power. Classical EPI results, like those introduced in~\cite{dePalmaEPI}, imply $e^{2H(X+Y)} \geq \tfrac{1}{2}(e^{2H(X)} + e^{2H(Y)})$, with equality when $ Y $ is Gaussian. Applying the entropy formula $ H(X) = \tfrac{1}{2} \log(2\pi e \sigma_X^2) $, this leads to a lower bound for the capacity: $ \tfrac{1}{2} \log(1 + \sigma_X^2/P) $, assuming $ N(Y) = e^{2H(Y)} $ and jammer power $ P $. Hence, for a sender power constraint $ E $, the achievable capacity becomes $ \frac{1}{2} \log\left(1 + \frac{E}{P}\right) $. 

Recent developments in quantum information theory have extended EPI-like results to quantum systems~\cite{koenigEntropyPower2013,dePalmaEPI}. However, as noted in~\cite{dePalmaEPI}, the quantum variant $ e^{2S(X \boxplus_\tau Y)} \geq \tau e^{2S(Y)} + (1 - \tau) 2^{2S(X)} $ generally fails to reach equality for Gaussian states with scaled covariance matrices. Moreover, when analyzing the capacity of $ \boxplus_\tau $ under thermal jamming with power $ P $, the known formula $ g(\tau E + (1 - \tau)P) - g((1 - \tau)P) $, where $g(x) = (x+1) \log(x+1) - x \log(x)$ is the Gordon function \cite{holevoBOOK}, suggests that a more appropriate quantum analogue would replace the exponential function with the inverse of $ g $, if one aims to reproduce the classical capacity expression using an entropy inequality.

The conjectured entropy photon number inequality (EPnI)~\cite{guhaEPnI}, which posits $ N(X \boxplus_\tau Y) \geq \tau N(Y) + (1 - \tau)N(X) $ with $ N = g^{-1} \circ S $, remains unproven. Nevertheless, a new form of this inequality, recently conjectured in~\cite{noetzel-isit2024}, enables us---assuming its validity---to establish an analytical capacity formula for the bosonic AVC $ \boxplus_\tau $, under the restriction to classical coherent jamming strategies with bounded energy.


\section{Notation}
Alphabets are boldfaced (e.g. $\mathbf X$) if finite and calligraphic (e.g. $\mathcal X$) if infinite. The set of probability measures on a set $\mathcal A$ is written $\mathcal P(\mathcal A)$, and the set of states on a Hilbert space $\mathcal H$ is $\mathcal P(\mathcal H)$.  The trace of an operator $A$ on $\mathcal H$ is $\tr(A)$, the scalar product of $x,y\in\mathcal H$ $\langle x,y\rangle$. The logarithm $\log$ is taken with respect to base two, and the entropy of $\rho\in\mathcal P(\mathcal H)$ is $S(\rho):=-\tr(\rho\log\rho)$. The binary entropy is $h:[0,1]\to[0,1]$. For a set $X$, $\mathcal C(X,\mathcal K)$ denotes classical-quantum channels mapping $x\in X$ to a quantum state. The Holevo quantity of a measure $\mu$ on $\mathcal P(\mathcal H)$ and quantum channel $\mathcal N$ is $\chi(\mu,\mathcal N)=S(\int \mu(x)\mathcal N(x)d\mu(x))-\int \mu(x)S(\mathcal N(x))d\mu(x)$. The one-norm on $\mathcal P(\mathcal H)$ is denoted as $\|\cdot\|_1$. For $a\in[0,1]$ we set $a':=1-a$.
The relative entropy of $\rho,\sigma\in\mathcal P(\mathcal H)$ is $D(\rho\|\sigma)$. For an operator $A$, $A_i:=\eins\otimes\ldots\otimes A\otimes\ldots\otimes\eins$ where $A$ is in the $i$-th position. The single-mode Fock space is $\mathcal F$, with corresponding sub-spaces $\mathcal F_N:=\mathrm{span}\{|0\rangle,\ldots,|N\rangle\}$ and projections $P_N:=\sum_{n=0}^{N}|n\rangle\langle n|$. The corresponding channel is $\boxminus_d(X):=P_dXP_d+\tr((\eins-P_d)X)|0\rangle\langle0|$.

We define the Hamiltonian $\mathbb H:=\sum_{n=0}^\infty n|n\rangle\langle n|$ and set $\mathcal P_{F}:=\{\rho:\tr(\rho\mathbb H)\leq F\}$. The displaced phase-randomized coherent states (DPHAV) are
$P_\beta(\alpha):=\tfrac{1}{2\pi}\int_0^{2\pi}|\alpha+e^{\mathbbm i\phi}\beta\rangle\langle \alpha+e^{\mathbbm i\phi}\beta|d\phi$. They are semi-classical, meaning that they have positive P-representation and arise from mixing a phase-randomized state $P_\beta(0)$ and a coherent state $|\alpha\rangle\langle\alpha|$ at a beam splitter \cite{poissonStatesAtBeamsplitter}. Let $\mathcal S_F$ be the set of PHAV states $\rho(\mu)=\int_0^\infty P_b d\mu(b)\in\mathcal P_F$ for which $\mu$ is sub-Gaussian \cite[Proposition 2.5.2]{Vershynin_2018} with constant $K_1\leq F$. Generally, for $\mu\in\mathcal P(\mathbb C)$, $\rho(\mu):=\int|\alpha\rangle\langle\alpha|d\mu(\alpha)$. For $k>1$ we define $\mathcal S_F^k$ as the set of states $\rho=\otimes_{i=1}^k|\alpha_i\rangle\langle\alpha_i|$ such that $\sum_i|\alpha_i|^2\leq k\cdot F$ for which the empirical distribution of $\alpha^n$ on $\mathbb C$ is sub-Gaussian with constant $K_1\leq F$. $\mathcal D_F$ denotes sub-Gaussian distributions on $\mathbb C$ with constant $K_1\leq F$. Finally, $S_N$ denotes the thermal state of mean photon number $N$ \cite{holevo-book}.
\section{Definitions}
We study a specific class of quantum arbitrarily varying and compound channels of practical relevance, where, mathematically speaking, communication takes place over a beam splitter. One of the input ports is used by the legal transmitter, and the other is used by the jammer. A legal receiver observes one fixed output port:
\begin{definition}[System Model]
    Let the beam splitter have fixed but arbitrary transmittivity $\tau$. Denote jammer quantum signals by $\sigma$ and transmitter signals by $\alpha\in\mathbb C$. Jammer states are restricted to having positive Glauber--Sudarshan P representation, i.e. $\sigma=\int |\alpha\rangle\langle\alpha|d\mu(\alpha)$. The channel acts as
    \begin{align}\nonumber
        \mathcal N_\sigma(\alpha)&=\int |\sqrt{\tau}\alpha+\sqrt{\tau'}\beta\rangle\langle\sqrt{\tau}\alpha+\sqrt{\tau'}\beta|d\mu(\beta)\\
        &= \int |\alpha\rangle\langle\alpha|\boxplus_\tau|\beta\rangle\langle\beta|d\mu(\beta) =\mathcal N(\alpha,\sigma)
    \end{align}
    The notation $\mathcal N$ and $\boxplus_\tau$ is used interchangeably. If the jammer is restricted to use inputs of the form $\rho^k=(\int\rho_xd\mu(x))^{\otimes k}$ where again $\tr\mathbb H\rho^k\leq k\cdot P$ holds, we speak of a \emph{compound} channel in what follows and use the symbol $\mathcal W$.
\end{definition}
\begin{definition}[$(k,\lambda)$ Code]\label{def:code}
    A $(k,\lambda)$ deterministic code $\mathcal C$ for the AVC $\mathcal N$ consists of a finite collection $\{x_m^k\}_{m=1}^M\subset\mathcal X^n$ of signals with corresponding input states $x_m^k:=(x_{m,1},\ldots,x_{m,k})$ satisfying $\sum_i|x_i|^2\leq E$ and a POVM $\{D_m\}_{m=1}^M$ on $\mathcal F^{\otimes k}$ for which the success probability 
    \begin{align}\label{def:p_s}
        p_s(\mathcal C):=\inf_{\sigma\in\mathcal S_P^k}\frac{1}{M}\sum_{m=1}^M\tr(D_m \mathcal N^{\otimes k}(x^k_m,\sigma))
    \end{align}
   satisfies $p_s(\mathcal C)\geq1-\lambda$. If the infimum in \eqref{def:p_s} is instead taken over the set $\mathrm{conv}\{\sigma^{\otimes k}:\sigma\in\mathcal S_P\}$ we use the symbol $\mathcal W$ and speak of a deterministic $(k,\lambda)$ code for the \emph{compound} channel $\mathcal W$. We then write $p_s(\mathcal C|\mathcal W)$ for the success probability.
    A $(k,\lambda)$ CR assisted code consists of a probability measure $\mu$ over the set of codes. Its success probability is defined as 
    \begin{align}\label{def:p_s-random-code}
        p_s(\mathcal C):=\inf_{\sigma\in\mathcal S_P^k}\int\sum_{m=1}^M\frac{\tr(D_m^\gamma \mathcal N^{\otimes k}(x^k_{\gamma,m},\sigma))}{M}d\mu(\gamma).
    \end{align}
\end{definition}
    We explicitly allow infinite amounts of CR to later use phase randomization together with permutations of the signals as CR.
\begin{definition}[Achievable Rates, Capacity]
    A rate $R\geq0$ is achievable for the classical-quantum arbitrarily varying channel $\mathcal N$ under state constraint $\mathcal R$ if there exists a sequence $(\mathcal C_n)_{k\in\mathbb N}$ of $(k,\lambda_n)$  codes, obeying the state constraint $\mathcal R$, such that both $\lambda_k\to0$ and $\limsup_{k\to\infty}\frac{1}{k}\log M_k\geq R$. 
    The message transmission capacity of $\mathcal N$ under average error criterion is defined as the supremum over all rates that are achievable for $\mathcal N$. It is denoted as $C(\mathcal N)$ for CR assisted codes and $C_d(\mathcal N)$ for deterministic codes. The capacity for the compound channel $\mathcal W$ is defined analogously.
\end{definition}

\section{Results} 
We first prove an entropic optimization formula for $\mathcal C(\mathcal N)$: 
\begin{theorem}\label{thm:main}
The CR-assisted capacity of $\mathcal N$ is given by 
\begin{align}\label{eq:capacity-formula-one}
    C(\mathcal N)=\sup_{\mu\in\mathcal D_E}\inf_{\sigma\in\mathcal S_P}\chi(\mu;\mathcal N_\sigma).
\end{align}
For every $\eps>0$ there exists a sequence of codes with rates $R_k\to C(\mathcal N)-\eps$ where for every $k>0$ the measure in \eqref{def:p_s-random-code} is discrete with support upper bounded by $k^2$.
\end{theorem}
The question arises whether a more explicit expression for \eqref{eq:capacity-formula-one} can be found. As it turns out, this question can be answered in the affirmative if a recently conjectured \cite{noetzel-isit2024}, new entropy power inequality turns out to be true. Let us define, for a density operator $X$ of a single bosonic mode and every $\lambda\in[0,1]$, the functions $R_\lambda(X):=g^{-1}\circ S(|0\rangle\langle0|\boxplus_\lambda X)$ and $L_\lambda(X):=g^{-1}\circ S(X\boxplus_\lambda|0\rangle\langle0|)$. The conjectured inequality \cite{noetzel-isit2024} states that for one-mode bosonic systems $X,Y$ and every $\tau\in[0,1]$ the following inequality holds:
    \begin{align}
        S(X\boxplus_\lambda Y)\geq g(L_\lambda(X) + R_\lambda(Y)). \tag{C} \label{conjecture}
    \end{align}
\begin{theorem}\label{thm:entropic-formula}
    If the conjectured inequality \eqref{conjecture} holds, then the capacity formula in Theorem \ref{thm:main} simplifies to $C(\mathcal N)=g(\tau E+\tau'P)-g(\tau'P)$.
\end{theorem}
Our Theorem \ref{thm:main} partially generalizes \cite{csiszarNarayan} to quantum systems. Unlike \cite{csiszarNarayan}, we cannot use a minimum distance decoder, since this concept lacks a suitable quantum analogue. Instead we use a technique similar to \cite{fullyQuantumAVC,quantumAVC}, using CR to transform the attack into a mixture of i.i.d. attacks, which can be handled efficiently by compound codes. Thereby \eqref{eq:capacity-formula-one} is proven in Subsection \ref{subsec:CR-capacity}. Difficulties emerging from infinite dimensions in our system are dealt with by applying (common) phase randomization. This trick transforms the attack into one composed of PHAV states. Since these are all mutually diagonal in the photon number basis, classical statistical methods can be used to provide a de-Finetti type of upper bound going beyond \cite{lancienWinter} by capturing the state constraint. We observe that all involved systems have effective dimension scaling as $\log k$ where $k$ is the number of channel uses. This graceful scaling allows us to benefit from \cite[Lemma 15]{winter2016tight}, so that \eqref{eq:capacity-formula-one} can be proven. 
In Subsection \ref{subsec:entropic-cr-formula} we prove \eqref{thm:entropic-formula} under the assumed validity of the recently conjectured entropy power inequality.

\section{Proof of Theorem \ref{thm:main}}    
\subsection{Common Randomness Assisted Capacity}\label{subsec:CR-capacity}
Assume the transmitter and receiver have a non-assisted code $\mathcal C$ available. For an arbitrary sequence of angles $\theta^k\in[0,2\pi)^k$ and any permutation $\pi\in S_k$, we set 
    \begin{align}\label{eqn:random-decoder}
        D'_{m,\theta^k,\pi}&:=U_\pi\big(\bigotimes_{i=1}^kV(\theta_i)\big)D_m\big(\bigotimes_{i=1}^kV(-\theta_i)\big)U_\pi^{-1},\\
        \rho'_{m,\theta^k,\pi}&:=U_\pi\big(\otimes_{i=1}^kV_{\theta_i}|\alpha_i\rangle\langle \alpha_i|V_{-\theta_i}\big)U_\pi^{-1}\label{eqn:random-encoder}
    \end{align}
    where $U_\pi$ is the usual representation of $\pi$ as a unitary on $\mathcal F^{\otimes k}$ and $V_\theta:=e^{\mathbbm i\theta \mathbb H}$ is a phase rotation by an angle $\theta$. Consider the random variable $X$ with uniformly random values $x$ in $[M]\times[0,2\pi)^k\times S_k$. Denote by $\hat{\mathcal{C}}$ a code where sender and receiver jointly apply phase rotations which they select independently between channel transmissions according to a uniform distribution. At the same time they apply permutations of the transmissions over the channel, again uniformly at random. We then obtain 
    \begin{align}
        p_s&(\hat{\mathcal{C}}) = 
            \inf_{\sigma^k\in\mathcal S_F^k}\frac{1}{M}\sum_m\tr D_{m}\mathcal N^{\otimes k}(\alpha^k_m\otimes \bar\sigma^k),
    \end{align}
    where for every $\sigma^k=\otimes_{i=1}^k\sigma_i$ the state $\bar\sigma^k$ is defined as
    \begin{align}
        \bar\sigma^k:=\frac{1}{k!}\sum_\pi U_\pi\big(\bigotimes_{i=1}^k\int_0^{2\pi}V(\theta)\sigma_{i}V(\theta)^{-1}d\theta\big) U_\pi^{-1}.
    \end{align}
    We now apply a sequence of steps to effectively transform such a permutation-invariant attack into a convex combination of i.i.d. attacks. For this, we would first like to employ the following Lemma:
    \begin{lemma}\label{lem:local-to-global-trace-bound}
        Let $\rho\in\mathcal S(\hr{k})$ satisfy $\min_{1\leq i\leq k}\tr Q(\tr_{\neq i}\rho)\geq1-\eps$ for some projection $Q$ on $\hr{1}$, where $\tr_{\neq i}$ denotes the partial trace over all systems except the $i$-th one. Then $\tr Q^{\otimes k}\rho\geq1-k\cdot\eps$.
    \end{lemma}
    We will apply Lemma \ref{lem:local-to-global-trace-bound} to our case as a method for bounding the error when utilizing techniques from the analysis of finite-dimensional quantum systems. This approach necessitates arguing first that a projection $Q$ such as the one described on Lemma \ref{lem:local-to-global-trace-bound} exists. This existence again will be proven based on bounds on the power of the sender and the jammer. In this regard, we note that the averaged jamming signal $\bar\sigma^k$ is by construction subject to the following conclusion:
   \begin{lemma}\label{lem:averaging-implies-locality-of-power-constraint}
        Let $\sigma\in\mathcal S^k_P$ and $\bar\sigma:=\tfrac{1}{k!}\sum_{\pi\in S_k}U_\pi\rho U_\pi^{-1}$. Then for each $i\in\{1,\ldots,k\}$ the reduced state $\tr_{\neq i}\bar\sigma$ is an element of $\mathcal S_P$.
   \end{lemma}
    For any fixed value $P>0$, we can now apply Lemma \ref{lem:averaging-implies-locality-of-power-constraint} together with the following bound in Lemma \ref{lem:effective-dimension-for-phase-randomized-coherent-state} to regulate the effective local dimension of $\bar\sigma^k$.
    \begin{lemma}\label{lem:effective-dimension-for-phase-randomized-coherent-state}
        Let $\rho\in\mathcal S_P$. If $N\geq 22\cdot\max\{2,|\alpha|^2\}$ then $\tr P_N\rho_\alpha\geq 1-2^{2-N\cdot\min\{2,\log2(e)/22\cdot K^2\}}$. 
    \end{lemma}
    We choose a constant $c_1$ such that with $N=\lfloor c_1(2+\min\{2,\log2(e)/22\cdot K^2\}\log k)\rfloor$ we achieve $\min_{\rho\in\mathcal S_P}\tr(P_N\rho)\geq1-k^{-t}$. Since $\tr_{\neq i}\bar\sigma^k\in\mathcal S_P$ by construction, Lemma \ref{lem:effective-dimension-for-phase-randomized-coherent-state} justifies the application of Lemma \ref{lem:local-to-global-trace-bound} to $\bar\sigma^k$ with $\epsilon=k^{-t}$ and $Q:=P_{N}$ and shows that the jamming input may be uniformly approximated by pruned states $\bar\sigma':= P\bar\sigma^k P$, where $P=\otimes_{i=1}^kP_N$. By the Gentle Operator Lemma \cite{wildeBOOK,Winter1999}
    \begin{align}
        \|P\sigma^k P - \sigma^k\|_1\leq2k^{-(t-1)/2}=:\xi_k, 
    \end{align}
    which implies $\|\rho^k\otimes P\sigma^k P - \rho^k\otimes \sigma^k\|_1\leq\xi_k$ and then by monotonicity of $\|\cdot\|_1$ the estimate $\|\mathcal N^{\otimes k}(\alpha^k\otimes P\sigma^k P - \alpha^k\otimes\sigma^k)\|_1\leq\xi_k$. The triangle inequality yields
    \begin{align}
        \tr D_m\mathcal N^{\otimes k}(\alpha^k_m\otimes\bar\sigma) \geq \tr D_m\mathcal N^{\otimes k}(\alpha^k_m\otimes\bar\sigma') - \xi_k
    \end{align}
    for all $m\in[M]$. This estimate shows that the use of common randomness in the sense of \eqref{eqn:random-decoder} and \eqref{eqn:random-encoder} effectively transforms the jamming strategy into one which may be described solely by permutation-invariant states which can locally be written in the form $\sum_{n=0}^N\lambda_n|n\rangle\langle n|$. However, a further step is needed to finally reduce the proof of Theorem \ref{thm:main} to the compound channel coding problem. This reduction finally allows us to apply a de-Finetti type of estimate, which we distribute over the following two statements: 
    \begin{lemma}\label{lem:concentration-for-av-distribution}
        Let $p^k:=\prod_{i=1}^kp_i$ for $p_1,\ldots,p_k\in\mathcal P([d])$. Define $\bar p^k:=\tfrac{1}{k!}\sum_{\pi\in S_k}\prod_{i=1}^kp_{\pi(i)}$ as well as $\bar p:=\tfrac{1}{k}\sum_{i=1}^kp_i$. Then
        \begin{align}
            \sum_{\|\bm-\bar p\|_1\leq\eps} \bar p^k(T_\bm)\geq1- C\cdot(2k)^d\cdot \exp\{-\eps^2\cdot k\},
        \end{align}
        where the sum is over empirical distributions satisfying for every $i\in[d]$ that $\bm(i)=\frac{m_i}{k}$ for $m_i\in\mathbb N$ and $T_m$ is the type set of $\mathbf m$: $T_{\mathbf m}:=\{x^k:N(i|x^k)=\mathbf m(i)\}$ with $N(i|x^k)$ being the number of times that the symbol $i$ appears in $x^k$.
    \end{lemma}
    Based on Lemma \ref{lem:concentration-for-av-distribution} we can conclude that the effective strategy of the jammer has a highly specific structure, which we can finally relate to the i.i.d. jamming strategies which we require for our intended application of compound codes to the given scenario via the following:
    \begin{lemma}\label{lem:flat-type-to-iid}
        Let $\bm$ be an empirical distribution so that for all $i\in[N]$ we have $\bm(i)=\tfrac{m_i}{k}$ for some $m_i\in\mathbb N$. Then 
        \begin{align}
            |T_\bm|^{-1}\eins_{T_\bm}\leq(2k)^N\bm^{\otimes k}.
        \end{align}
    \end{lemma}
    Lemma \ref{lem:flat-type-to-iid} is an immediate consequence of the method of types. We provide a sketch of the proof in the appendix. 
    
    We emphasize that $\bar\sigma\in\mathcal S_P$ is diagonal in the number state basis and that every $\bm$ can be interpreted as the quantum state $|\bm_1\rangle\langle\bm_1|\otimes\ldots\otimes|\bm_k\rangle\langle\bm_k|$, which we will also denote as $\bm$ the context is unambiguous. 
    
    We now proceed to finalizing our de-Finetti type of estimate by setting $B_\sigma:=\{\bm:\|\bm-\bar p\|_1\leq\eps\}$:
    \begin{align}
        &p_s(\hat{\mathcal{C}}) \geq 1-\inf_{\sigma\in\mathcal S_P^k}\sum_{m=1}^M\frac{\tr(D_m \mathcal N^{\otimes k}(\alpha^k_m\otimes\bar\sigma')}{M}- \xi_k\\
        &\geq 1-\inf_{\sigma\in\mathcal S_P}\frac{(2k)^N}{M}\sum_{m=1}^M\sum_{\bm\in B_\sigma}\lambda_\bm\tr D_m \mathcal N^{\otimes k}(\alpha^k_m\otimes \bm^{\otimes k})\nonumber\\
        &\qquad\qquad - C\cdot(2k)^N\cdot \exp\{-\eps^2\cdot k\} - \xi_k\\
         &\geq 1-\inf_{\bm\in \big(\mathcal S_P\big)_\eps}\frac{(2k)^N}{M}\sum_{m=1}^M\tr D_m \mathcal N^{\otimes k}(\alpha^k_m\otimes \bm^{\otimes k})\nonumber\\
        &\qquad\qquad - C\cdot(2k)^N\cdot \exp\{-\eps^2\cdot k\} - \xi_k\label{eqn:av-decoding-error}
    \end{align}
    where $(\mathcal S_P)_\eps:=\{\bm\in\mathcal P([N]):\min_{\sigma\in\mathcal S_P}\|\sigma-\bm\|_1\leq\eps\}$ and $\lambda_\bm$ are positive weights summing up to one.
    To derive a lower bound on the error probability of the CR-assisted code $\hat{\mathcal{C}}$, it thus suffices to prove a sufficiently good lower bound on the success probability of $\mathcal C$ when used for the $\emph{compound}$ channel $\mathcal W$. The term ``sufficiently good'' here means that the error $\eps_k$ should satisfy $\eps_k\cdot 2^{(\log k)^2}\to0$.
    
    Here we make use of the techniques from \cite{cnr21}, which again are based on the work \cite{bbjn2012}. We apply these techniques to our case by first fixing some arbitrary $d>0$. For arbitrary complex numbers $\alpha\in\mathbb C$ with, define the classical-quantum channels
    \begin{align}\label{def:finite-dim-approximation}
        \boxminus_{d,\sigma}(\alpha):=\boxminus_d\circ\boxplus_\tau(|\alpha\rangle\langle\alpha|\otimes\sigma).
    \end{align} 
    We specify $\mathcal C_k$ to be a $(k,\lambda)$ code for the compound channel $\{\boxminus_{d,\bm}\}_{\bm\in(\mathcal S_P)_\eps}$. We aim to use the following Lemma:
    \begin{lemma}[{\cite[Lemma 1]{bbjn2012}}\label{lem:compound-code-1}]
        Let $\{\mathcal N_s\}_{s\in\mathbf S}$ be a compound channel where $\mathcal N_s\in \mathcal C(\mathcal H,\mathcal K)$ for finite-dimensional Hilbert spaces $\mathcal H$ and $\mathcal K$ and let $p\in\mathcal P(\mathbf X)$, where $\{|x\rangle\}_{x\in\mathbf X}$ denotes an orthonormal basis of $\mathcal H$. Define $\mathbf{p}:=\sum_xp(x)|e_x\rangle\langle e_x|$, 
        \begin{align}
            \rho_{s,k}&:=\frac{1}{|\mathbf S|}\sum_s\sum_{x^k}p^{\otimes k}(x^k)|e_{x^k}\rangle\langle e_{x^k}|\otimes \mathcal N_s^{\otimes k}(x^k)\\
            \sigma_{s,k}&:=\frac{1}{|\mathbf S|}\sum_s\mathbf{p}^{\otimes k}\otimes \sum_{x^k}p^{\otimes k}(x^k)\mathcal N_s^{\otimes k}(x^k).
        \end{align}
        If there is a projector $q_k$ such that
        \begin{align}
            tr(q_k\rho_{k})&\geq1-\lambda,\qquad
            tr(q_k\sigma_{k})\leq2^{-k\cdot a}
        \end{align}
        then for any $0<\gamma\leq a$ there is a $(k,|\mathbf S|(2\cdot \lambda + 4\cdot2^{-{k}\gamma}))$ code with $M=\lceil2^{k(a-\gamma)}\rceil$ for the compound channel $\{\mathcal N_s\}_{s\in\mathbf S}$.
    \end{lemma}
    \begin{lemma}\label{lem:compound-code-2}
        For every $\delta>0$ and $p\in\mathcal P(\mathcal X)$ there is a $\tilde c$ such that, for large enough $k$, there is a projector $q_k$ satisfying
        \begin{align}
            tr(q_k\rho_{k})&\geq1-|\mathbf S|2^{-k\cdot\tilde c},\ \tr(q_k\sigma_{k})&\leq2^{-k\cdot(a-\delta)},
        \end{align}
        where $a:=\min_{s\in\mathbf S}D(\rho_{s,1}\|\mathbf p\otimes \sigma_{s,1}) = \min_{s\in\mathbf S}\chi(p,\mathcal N_s) $.
    \end{lemma}
    The set $(\mathcal S_P)_\eps$ is a subset of $\mathcal P([N])$ and therefore above results do not apply directly. We can however approximate this set with a finite one to sufficient accuracy. We pick a regular grid $\mathbf \Delta_L$ of size $L$ on $\mathbb C$, which we adjust later. 

    We are now ready to invoke Lemma \ref{lem:compound-code-1} and Lemma \ref{lem:compound-code-2}. As observed already in \cite{cnr21}, there is a critical point in the proof where a parameter $w(k)$ is introduced. With our choice of dimension, it holds $w(k) := \tfrac{t^2\log^2k}{k}\log(k + 1)$ (compare \cite[Eq. 63]{bbjn2012}). The proof in \cite{bbjn2012} further rests on the following estimate \cite[Eq. 74]{bbjn2012}:
    \begin{align}
            f'_{k,\nu}(0)\leq-\tfrac{\delta}{2}+\tfrac{1}{k}\log (2k)^N<0,
    \end{align}
    where $(2k)^N$ is the number of possible jammer strategies. Apparently this inequality is eventually true due to our choice $N\sim\log k$. 
    
    It follows for every distribution $p\in\mathcal P(\Delta_L)$ the existence of a sequence $(\mathcal C_n)_{n\in\mathbb N}$ of codes with rates $(R_n)_{n\in\mathbb N}$ 
    satisfying $\lim_{n\to\infty}R_n= \inf_{\sigma\in(\calS_P)_\eps}\chi(p;\boxminus_{d,\bm})-\delta$. The proof of \cite[Lemma 1]{bbjn2012} can easily be modified to work with the pruned distributions $p'(x^k):=c\cdot p^{\otimes k}(x^k)$ for $\|\tfrac{1}{k}N(\cdot|x^k)-p\|_1\leq\delta$ and $p'(x^k)=0$, else where $c$ is a suitable normalization constant. This way, one obtains codes with controlled average energy $\tfrac{1}{k}\sum_i|\alpha_{m,i}|^2\leq E$ for every code-word $m\in[M]$. It follows that 
    \begin{align}
        C(\mathcal N)\geq \inf_{\sigma\in\calS_P}\chi(p;\boxminus_{d,\sigma})
    \end{align}
    for every $p\in\mathcal P(\Delta_L)$ satisfying $\sum_{\alpha\in\Delta_L}p(c)|\alpha|^2< E$. Moreover, since the error of our code decays as $\sim k^{-t'}$, an application of the Markov inequality shows that already $k^2$ randomly drawn codes with random permutations and phase changes suffice. 
    
    To finally show \eqref{eq:capacity-formula-one} we use the continuity of entropy on the sets $\mathcal P_{E}$ which had already been mentioned in \cite{wehrl1978}:
    \begin{lemma}[{\cite[Gibbs Hypothesis and Lemma 15]{winter2016tight}}]\label{lem:gibbs-hypothesis}
            If $tr\left(\exp\{-\beta \mathbb{H}\}\right)<\infty$ for all $\beta>0$ and $\max\{\tr(\rho \mathbb{H}),\tr(\sigma \mathbb{H})\}\leq E$ for some $E>0$ then $\|\rho-\sigma\|_1\leq\eps$ implies $|S(\rho) - S(\sigma)|<\eps C(\eps,\mathbb{H},E) + h(\eps)$ for a function $C$ satisfying $\lim_{\eps\to0}\eps C(\eps,\mathbb{H},E)=0$.
    \end{lemma} 
    Let $\mu$ be a mean-zero Gaussian distribution on $\mathbb C$ with variance equal to $E'<E$. 
    Choose $\Delta_L$ as the points of a regular square grid approximating every point on the circle $\{\alpha:|\alpha|^2\leq E''\}$ to accuracy $\epsilon$. The parameters $E'',\epsilon,L$ are related via $L\leq (4E'/\epsilon)^2$. To each point $\alpha\in\Delta_L$ in the grid we assign the surrounding box 
    \begin{align}
        \boxdot_c:=\left\{c':\begin{array}{ll}\tfrac{-\epsilon}{2}<\Re{(c-c')}\leq\tfrac{\epsilon}{2}\\ \tfrac{-\epsilon}{2}<\Im{(c-c')}\leq\tfrac{\epsilon}{2}\end{array}\right\}
    \end{align}
    and the number $\mu_c = \kappa^{-1}\cdot \mu(\boxdot_c)$ where $\kappa:=\sum_{c\in\Delta_L}\mu(\boxdot_C)$. Then $p(c):=\mu_c$ defines a probability distribution on $\Delta_L$. Further, we let $E'':=\epsilon^{-1}$. Then
    \begin{align}        
        \lim_{\epsilon\to0}\|\sum_{c\in\Delta_L}p(c)|c\rangle\langle c| - \int|\alpha\rangle\langle\alpha|d\mu(\alpha)\|_1=0
    \end{align}    
    and at the same time
    \begin{align}
        \lim_{\epsilon\to0}\max_{c\in\Delta_L}\max_{\alpha\in \boxdot_c} \| |c\rangle\langle c| - |\alpha\rangle\langle\alpha|\|_1 = 0.
    \end{align}
    Due to the monotonicity of the trace under CPTP maps, the specific form of $\boxminus_d$ which implies $\tr\mathbb H\boxminus_{d,\sigma}(\rho)\leq\tr\mathbb H\rho\boxplus_\tau\sigma$ and Lemma \ref{lem:gibbs-hypothesis} it follows that 
    \begin{align}        
        S(\mathcal N_\sigma(\sum_{\gamma\in\bGamma}p(\gamma)|\gamma\rangle\langle\gamma|))\to S(\mathcal N_\sigma(\int|\alpha\rangle\langle\alpha|d\mu(\alpha))
    \end{align}
    and in addition 
    \begin{align}
        \max_{c\in\Delta_L}\max_{\alpha\in \boxdot_c} \| S(\mathcal N_\sigma(|c\rangle\langle c|)) - S(\mathcal N_\sigma(|\alpha\rangle\langle\alpha|))\|_1\to0.
    \end{align}
    Finally, this implies for every $\epsilon'>0$ the existence of $\epsilon>0$ and a corresponding $L\sim\epsilon^{-4}$ such that
    \begin{align}
        \max_{p\in\Delta_L}\inf_{\sigma\in\calS_P}\chi(p;\hat{\mathcal{N}}_\sigma) \to \max_{\mu\in\mathcal D_E}\min_{\sigma\in\mathcal S_P}\chi(\mu;\mathcal{N}_\sigma)-\epsilon'
    \end{align}
    and thereby Theorem \ref{thm:main}. We note that the energy constraints are obeyed throughout by definition of our finite-dimensional approximation \eqref{def:finite-dim-approximation}. The converse follows by assuming a jamming strategy tailored to \eqref{eq:capacity-formula-one}, where the jammer sends the jamming signal $S_P^{\otimes k}$.

    \section{Proof of Theorem \ref{thm:entropic-formula}\label{subsec:entropic-cr-formula}}
    We will closely follow the strategy that was used in \cite{noetzel-isit2024}. To show that the conjectured inequality $S(X\boxplus_\lambda Y)\geq g(L_\lambda(X) + R_\lambda(Y))$ \cite{noetzel-isit2024} implies that \eqref{eq:capacity-formula-one} in fact equals $g(\tau E+\tau'P)-g(\tau'P)$ we first observe that, obviously, 
        \begin{align}
            C(\mathcal N)\geq\min_{\sigma\in\mathcal S_P}\chi(S_E;\boxplus_\tau\sigma).
        \end{align}
        Further we observe that 
        \begin{align}
            \int p(x)S(|x\rangle\langle x|\boxplus_\tau\sigma)dx
                &= S(|0\rangle\langle 0|\boxplus_\tau\sigma)= g(R_\tau(\sigma)).
        \end{align}
        We then invoke the inequality, arriving at the lower bound
        \begin{align}
            C&(\mathcal N)\geq \min_{\sigma\in\mathcal S_P}\big(S(S_E\boxplus_\tau\sigma)) - S(|0\rangle\langle 0|\boxplus_\tau\sigma)\big)\\
            &\geq\min_{\sigma\in\mathcal S_P}\big(g(L_\tau(S_E)+R_\tau(\sigma)) - g(R_\tau(\sigma)) \big).
        \end{align}
        We note that $g$ is strictly increasing \cite{guhaEPnI}, and thus its inverse is strictly increasing as well. Further $S_P$ maximizes $S(\sigma)$ on $\mathcal S_P$. Therefore $R_\tau(\sigma)$ takes the maximum value $R_\tau(S_P)=g^{-1}\circ S(|0\rangle\langle0|\boxplus_\tau S_P)=(1-\tau)P$. It follows that $C(\boxplus_\tau)=g(\tau E+\tau' P)-g(\tau' P)$.

    \begin{section}{Conclusion}
        We have derived a capacity formula for a quantum AVC of practical interest, thereby extending the earlier results \cite{csiszarNarayan} on classical AVCs and moving the investigations started in \cite{ahlswede-blinovsky,quantumAVC} towards bosonic systems. We utilized a recently conjectured quantum entropy power inequality. Conditioned on its validity, we have proven an analytical formula for the CR-assisted data transmission capacity of the AVC $\boxplus_\tau$. We note that conditions such as $K_1\leq F$ in our definition of the set of admissible jamming strategies $\mathcal S_F^k$ and, in general, the introduction of the constraint of ``sub-Gaussianity'' which we imposed on the jammer is a somewhat technical requirement. We hope that future work will improve on this restriction by developing and using novel tools.

        We assume that attempts to derandomize our code constructions will however benefit from the ``sub-Gaussianity'' state constraints imposed on the jammer in our model: The straightforward use of Ahlswede's elimination technique~\cite{ahlswede-elimination} is usually not the way to go in settings with power constraints.

    \end{section}

\newpage
\newpage

\bibliographystyle{IEEEtran}
\bibliography{bib}

\newpage 

\section*{Appendix}
        We now provide the proofs of the lemmas.

    \begin{IEEEproof}[Proof of Lemma \ref{lem:local-to-global-trace-bound}]
        It holds $\eins = Q+Q^\perp$ for some orthogonal projection $Q^\perp$. Let $Q^{(i)}:=Q^{\otimes i}\otimes\eins^{k-i}$ and $Q_i:=\eins\otimes\ldots\otimes Q\otimes\ldots\otimes\eins$ where $Q$ is in the $i$-th position. Then we have $Q_iQ^{(i-1)}=Q^{(i-1)}Q_i=Q^{(i)}$ for every $i\in\{1,\ldots,k\}$. Let the estimate $\tr Q^{(i-1)}\rho)\geq1-i\cdot\eps$ hold for some $i$. Then 
    \begin{align}
        \tr&(Q^{(i)}\rho) = \tr(Q_i Q^{(i-1)}\rho)\\
            &= \tr((\eins-Q^\perp)_i Q^{(i-1)}\rho)\\
            &\geq 1-(i-1)\eps - \tr(Q^\perp_i Q^{(i-1)}\rho)\\
            &\geq 1-(i-1)\eps - \tr(Q^\perp_i\rho)\\
            &= 1-(i-1)\eps - \tr(Q^\perp\tr_{\neq i}\rho)\\
            &= 1-(i-1)\eps - (1-\tr(Q\tr_{\neq i}\rho))\\
            &\geq 1-(i-1)\eps - (1-1+\eps)\\
            &= 1-i\cdot\eps .
    \end{align}
    Since the statement holds for $i=1$ by assumption, the result is proven by induction.
    \end{IEEEproof} 
    \begin{IEEEproof}[Proof of Lemma \ref{lem:averaging-implies-locality-of-power-constraint}]
        Let us first observe that $\rho$ has exactly the same energy as $\bar\rho$:
        \begin{align}
            \tr&\textstyle\sum_{i=1}^kH_i\bar\rho = \tr\sum_{i=1}^kH_i\tfrac{1}{n!}\sum_{\pi\in S_n}\pi\rho\pi^{-1}\\
                &= \textstyle\tfrac{1}{n!}\sum_{\pi\in S_n}\tr\sum_{i=1}^kH_{\pi(i)}\rho\\
                &= \textstyle\tfrac{1}{n!}\sum_{\pi\in S_n}\tr\sum_{i=1}^kH_{i}\rho\\
                &= \textstyle\tr\sum_{i=1}^kH_{i}\rho.
        \end{align}
        Due to permutation-invariance of $\bar\rho$ it holds that $\tr_{\neq i}\bar\rho=\tr_{\neq j}\bar\rho$ for all $i,j\in\{1,\ldots,k\}$. Therefore $\tr\mathbb H\tr_{\neq i}\bar\rho = \tr_{\neq j}\bar\rho$ for all $i,j\in\{1,\ldots,k\}$. It follows that 
        \begin{align}
            E&\geq\tr\textstyle\sum_{i=1}^k\mathbb H_i\bar\rho\\
                &= \tr\textstyle\sum_{i=1}^k\mathbb H_i\bar\rho\\
                &= \textstyle\sum_{i=1}^k\tr\mathbb  H \tr_{\neq i}\bar\rho\\
                &= k\cdot \tr \mathbb H \tr_{\neq 1}\bar\rho,
        \end{align}
        proving the desired estimate. We proceed with the second estimate. Assume $|\{i:|\beta_i|^2\geq x^2\}|\leq k\cdot\exp\{-x^2/K_1\}$ holds, so that the empirical distribution of the jamming sequence $\beta^n$ is indeed sub-Gaussian with constant $K_1\leq F$. Letting $\mu'(A):=|\{i:\beta_i\in A\}|/k$ denote the empirical distribution of $\{\beta_i\}_{i=1}^k$ on $\mathbb R_+$ we then have
        \begin{align}
            \mu'(\{b:b^2\geq x^2\}) &= \tfrac{|\{i:|\beta_i|^2\geq x^2\}|}{k}\leq e^{-\tfrac{x^2}{K_1}}.
        \end{align}
        We can now extend $\mu'$ to $\mathbb C$ in the natural way by setting $\mu(A)=\tfrac{1}{k}\sum_{i=1}^k\tfrac{1}{2\pi}\int_0^{2\pi}\chi_A(e^{\mathbbm i\phi}\beta_i)d\phi$. Then
        \begin{align}
            \mu(\{\beta:|\beta|^2\geq x^2\}) &= \tfrac{1}{k}\sum_{i=1}^k\tfrac{1}{2\pi}\int_0^{2\pi}\chi_{\{\beta:|\beta|^2\geq x^2\}}(e^{\mathbbm{i}\phi}\beta_i)d\phi\nonumber\\
            &=\tfrac{1}{k}\sum_{i=1}^k\chi_{\{\beta:|\beta|^2\geq x^2\}}(\beta_i)\\
            &= \tfrac{1}{k}|\{i:|\beta_i|^2\geq x\}\\
            &\leq\exp\{-x^2/K_1\}.
        \end{align}
    \end{IEEEproof}
    \begin{IEEEproof}[Proof of Lemma \ref{lem:effective-dimension-for-phase-randomized-coherent-state}]
    For $P_N:=\sum_{n=0}^{N-1}|n\rangle\langle n|$ and $\rho= P_\beta(0)$ we know from \cite[Eq. (III.V)]{noetzelInfinite}
        \begin{align}\label{eqn:trace-lower-bound}
            \tr& P_N\rho \geq 1 - 4^{-N},
        \end{align}
        whenever $N\geq 22\cdot\max\{2,|\beta|^2\}$. Most importantly, for $N\geq44$ (which we assume to hold from now on) the inequality $N\geq22\cdot|\beta|^2$ must be satisfied for \eqref{eqn:trace-lower-bound} to hold. We recall the definition of sub-Gaussianity \cite[Theorem 2.5.2]{Vershynin_2018} of a random variable $X$ which requires $\mathbb P(|X|\geq t)\leq 2\exp(-t^2/K^2)$.
        Taking now a sub-Gaussian distribution $\mu$ with any constant $K$, and setting $\kappa=1/22$, we obtain
        \begin{align}
            \tr\rho P_N 
                &\geq \int_{B_{\sqrt{\kappa\cdot N}}}\tr[P_N\rho_\beta] d\mu(\beta)\\
                &\geq \int_{B_{\sqrt{\kappa\cdot N}}}(1-4^{-N})d\mu(\beta)
        \end{align}
        where the last inequality holds due to our choice of $\kappa$ and the bound \eqref{eqn:trace-lower-bound} for all $N\geq 44$. It remains to utilize sub-Gaussianity of $\mu$:
        \begin{align}
            \tr\rho P_N 
                &\geq (1-4^{-N})\mu(B_{\sqrt{\kappa\cdot N}})\\
                &\geq(1-4^{-N})(1-2\exp(-\kappa\cdot N/K^2))\\
                &=(1-2^{-2N})(1-2^{1-\kappa\log2(e)\cdot N/K^2})\\
                &\geq 1-2^{2-N\cdot\min\{2,\log2(e)/22\cdot K^2\}}.
        \end{align}
    \end{IEEEproof}
    \begin{IEEEproof}[Proof of Lemma \ref{lem:concentration-for-av-distribution}]
        Consider a fixed but arbitrary $n\in[d]$ and let the function $F:[d]^k\to\mathbb N$ be defined as $F(n^k):=N(n|n^k)$. If we modify $n^k$ in exactly one position to obtain a new string $\tilde n^k$, then $|F(n^k)-F(\tilde n^k)|\leq1$. By using Mc Diarmid's inequality, we thus have 
        \begin{align}
            p^k(\{n^k:F(n^k)-\mathbb E_{p^k}(F)|\geq\epsilon\})\leq C\cdot e^{-\eps^2/k}, 
        \end{align} 
        where we use the distribution $p^k:=\prod_{i=1}^kp_i$. 
        The expectation of $F$ is calculated as
        \begin{align}
            \mathbb E_{p^k}(F) &= \sum_{n^k}p^k(n^k)F(n^k)\\
                &= \sum_{n^k}p^k(n^k)N(n|n^k)\\
                &= \sum_{n^k}p^k(n^k)\sum_{i=1}^k\delta(n,n_i)\\
                &= k\cdot\bar p(n).
        \end{align}
        and due to the power constraint it satisfies $\mathbb ES^k\leq E\cdot k$. Therefore, for every $n\in[d]$, 
        \begin{align}
             p^k(\{n^k:F(n^k)-k\cdot \bar p(n)|\geq\epsilon\})\leq C\cdot e^{-\eps^2\cdot k}.
        \end{align}
        There are $(d+1)^k$ different types. Thus the set $A_k:=\{n^k:\|N(\cdot|n^k)-\bar p\|_1\geq \eps\cdot k\}$ satisfies
        \begin{align}\label{eqn:probability-estimate-for-A_k}
            p^k(A_k)\leq C\cdot(2d)^k\cdot \exp\{-\eps^2\cdot k\}.
        \end{align}
        Finally since $A_k$ is permutation-invariant, property \eqref{eqn:probability-estimate-for-A_k} holds for every permutation $\pi(p^k):=\prod_{i=1}^kp_{\pi(i)}$. Therefore we finally have
        \begin{align}
            \bar p^k(A_k)\leq C\cdot(2d)^k\cdot \exp\{-\eps^2\cdot k\}.
        \end{align}
    \end{IEEEproof}
    \begin{IEEEproof}[Proof of Lemma \ref{lem:flat-type-to-iid}]
        By \cite[Lemma 2.3]{csiszar-koerner-book} it holds 
        \begin{align}\label{eqn:size-bounds-for-typical-sets}
            2^{k\cdot H(\mathbf m)}\geq|T_{n^k}|\geq(1+k)^{-d}\cdot 2^{k\cdot H(\mathbf m)}.
        \end{align}
        For every $n^k\in T_\bm$ we have $N(n|n^k)=k\cdot \bm(n)$ and thus
        \begin{align}
            \bm^{\otimes k}(n^k) 
                &= \prod_{i=0}^dt(n)^{k\cdot N(n|n^k)}\\
                &=  2^{k\sum_{n=0}^dt(n)\log t(n)}\\
                &= 2^{-k\cdot H(\bm)}.
        \end{align}
        It follows that we have for every $n^k\in T_\bm$ the relation
        \begin{align}
            \pi_t(n^k) 
                &= |T_t|^{-1}\\
                &\leq (2k)^d\cdot 2^{k\cdot H(\bm)}\\
                &= (2k)^d\bm^{\otimes k}(n^k).
        \end{align}
        Since $\pi_\bm(n^k)=0$ for $n^k\notin T_\bm$, the result follows.
    \end{IEEEproof}

\end{document}